\documentclass[superscriptaddress,twocolumn,prl,showpacs,floatfix]{revtex4}
\usepackage[dvips]{graphicx}
\usepackage{amsmath,amsfonts,amssymb,bm,ulem}
\usepackage{graphicx}
\begin{document}

\title{Magnetoresistance in Dilute $p$-Si/SiGe in Parallel and Tilted Magnetic Fields}

\author{I.L.~Drichko}
\affiliation{A.F. Ioffe Physico-Technical Institute of Russian
Academy of Sciences, 194021 St. Petersburg, Russia}
\author{I.Yu.~Smirnov}
\email{ivan.smirnov&mail.ioffe.ru} \affiliation{A.F. Ioffe
Physico-Technical Institute of Russian Academy of Sciences, 194021
St. Petersburg, Russia}
\author{A.V.~Suslov}
\affiliation{National High Magnetic Field Laboratory, Tallahassee,
FL 32310, USA}
\author{O.A.~Mironov}
\affiliation{Warwick SEMINANO R$\&$D Centre, University of Warwick
Science Park, Coventry CV4 7EZ, UK}
\author{D.R. Leadley}
\affiliation{Department of Physics, University of Warwick, Coventry,
CV4 7AL, UK}

\begin{abstract}
We report the results of an experimental study of the
magnetoresistance  $\rho_{xx}$ and  $\rho_{xy}$ in two samples of
$p$-Si/SiGe with low carrier concentrations
$p$=8.2$\times$10$^{10}$\,cm$^{-2}$ and
$p$=2$\times$10$^{11}$\,cm$^{-2}$. The research was performed in the
temperature range of 0.3-2 K and in the magnetic fields of up to 18
T, parallel or tilted with respect to the two-dimensional (2D)
channel plane. The large in-plane magnetoresistance can be explained
by the influence of the \textit{in-plane} magnetic field on the
orbital motion of the charge carriers in the quasi-2D system.  The
measurements of  $\rho_{xx}$ and  $\rho_{xy}$ in the tilted magnetic
field showed that the anomaly in $\rho_{xx}$, observed at filling
factor $\nu$=3/2 is practically nonexistent in the conductivity
$\sigma_{xx}$. The anomaly in $\sigma_{xx}$ at $\nu$=2 might be
explained by overlapping of the levels with different spins
0$\uparrow$ and 1$\downarrow$  when the tilt angle of the applied
magnetic field is changed. The dependence of g-factor
$g^*(\Theta)/g^*(0^0)$ on the tilt angle $\Theta$ was determined.
\end{abstract}

 \maketitle

\section{Introduction}
In studied $p$-Si/Si$_{1-x}$Ge$_x$/Si the 30 nm wide asymmetrical
quantum well is positioned in the layer of strained
Si$_{1-x}$Ge$_x$, so threefold degenerated (not considering a spin)
SiGe valence band is split into 3 subbands via strong spin-orbit
interaction and a strain. Charge carriers are the heavy holes,
related to the band which is formed from the atomic states with
quantum numbers $L$=1, $S$=1/2, and $J$=3/2. As the result, there is
a strong anisotropy of g-factor: $g^*_{\perp} \cong$ 4.5 when the
magnetic field is perpendicular to the plane of the quantum well and
$g^*_{\parallel} \cong$ 0 when the magnetic field is oriented in the
plane of the well~\cite{1}.

One of the interesting phenomena observed in this object was the
discovery of the so-called "reentrant Metal-to-Insulator transition"
in a magnetic field at filling factor $\nu$=3/2~\cite{2,3,4,5}. In
Ref.\cite{2} that anomaly was attributed to the presence of
long-range potential fluctuations with amplitude comparable with the
Fermi energy in this material. However, the author of~\cite{3,4,5}
explained those magnetoresistance anomalies by crossing of Landau
levels with different spin directions 0$\uparrow$ and 1$\downarrow$
when the magnetic field increases. In the present work the studies
of magnetoresistance and Hall effect were conducted in tilted
magnetic fields to determine the dependence of $g$-factor on the
tilt angle and the possible causes of the anomalies in
magnetoresistance and conductivity at filling factor $\nu$=2 in the
sample $p$-Si/SiGe with $p$=2$\times$10$^{11}$\,cm$^{-2}$.

\section{Experimental Results and Discussion}
We report the results of an experimental study of the
magnetoresistance  $\rho_{xx}$ in two samples of $p$-Si/SiGe/Si with
low carrier concentrations $p$=8.2$\times$10$^{10}$\,cm$^{-2}$ and
$p$=2$\times$10$^{11}$\,cm$^{-2}$. The research was performed in the
temperature range of 0.3-2 K in the magnetic fields of up to 18 T,
parallel to the two-dimensional (2D) channel plane\cite{6} at two
orientations of the in-plane magnetic field $B_{\parallel}$  against
the current $I$: $B_{\parallel} \perp I$ and $B_{\parallel}
\parallel I$. In the sample with the lowest density in the magnetic
field range of 0-7.2 T the temperature dependence of $\rho_{xx}$
demonstrates the metallic characteristics ($d\rho_{xx}/dT>$0).
However, at $B_{\parallel}$ =7.2 T the derivative $d \rho_{xx}/dT$
reverses the sign. Moreover, the resistance depends on the current
orientation with respect to the in-plane magnetic field. At
$B_{\parallel} \cong$ 13 T there is a transition from the dependence
$\ln(\Delta\rho_{xx} / \rho_{0})\propto B_{\parallel}^2$ to the
dependence $\ln(\Delta\rho_{xx} / \rho_{0})\propto B_{\parallel}$.
The observed effects can be explained by the influence of the
in-plane magnetic field on the orbital motion of the charge carriers
in the quasi-2D system\cite{7}. This result confirms that in the
in-plane magnetic field $g^*\approx$0.
\begin{figure}
\includegraphics[width=70mm]{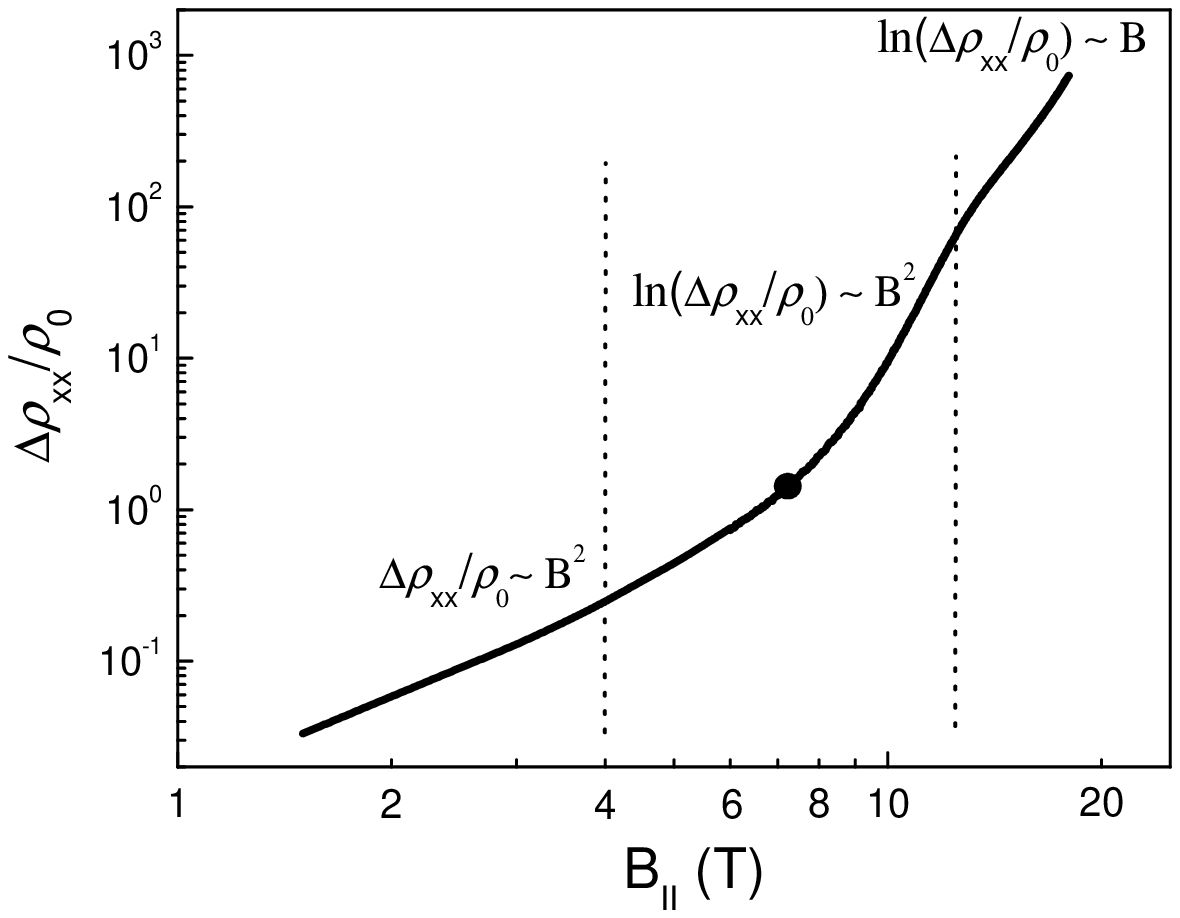}
\caption{Dependence of $\Delta \rho_{xx} / \rho_{0}$ on
$B_{\parallel}$ at $T$=0.3 K for the sample with
$p$=8.2$\times$10$^{10}$\,cm$^{-2}$. $B_{\parallel} \perp I$.}
\label{fig:sidecaption}
\end{figure}

\begin{figure}[h]
\includegraphics[width=7.8 cm,clip=]{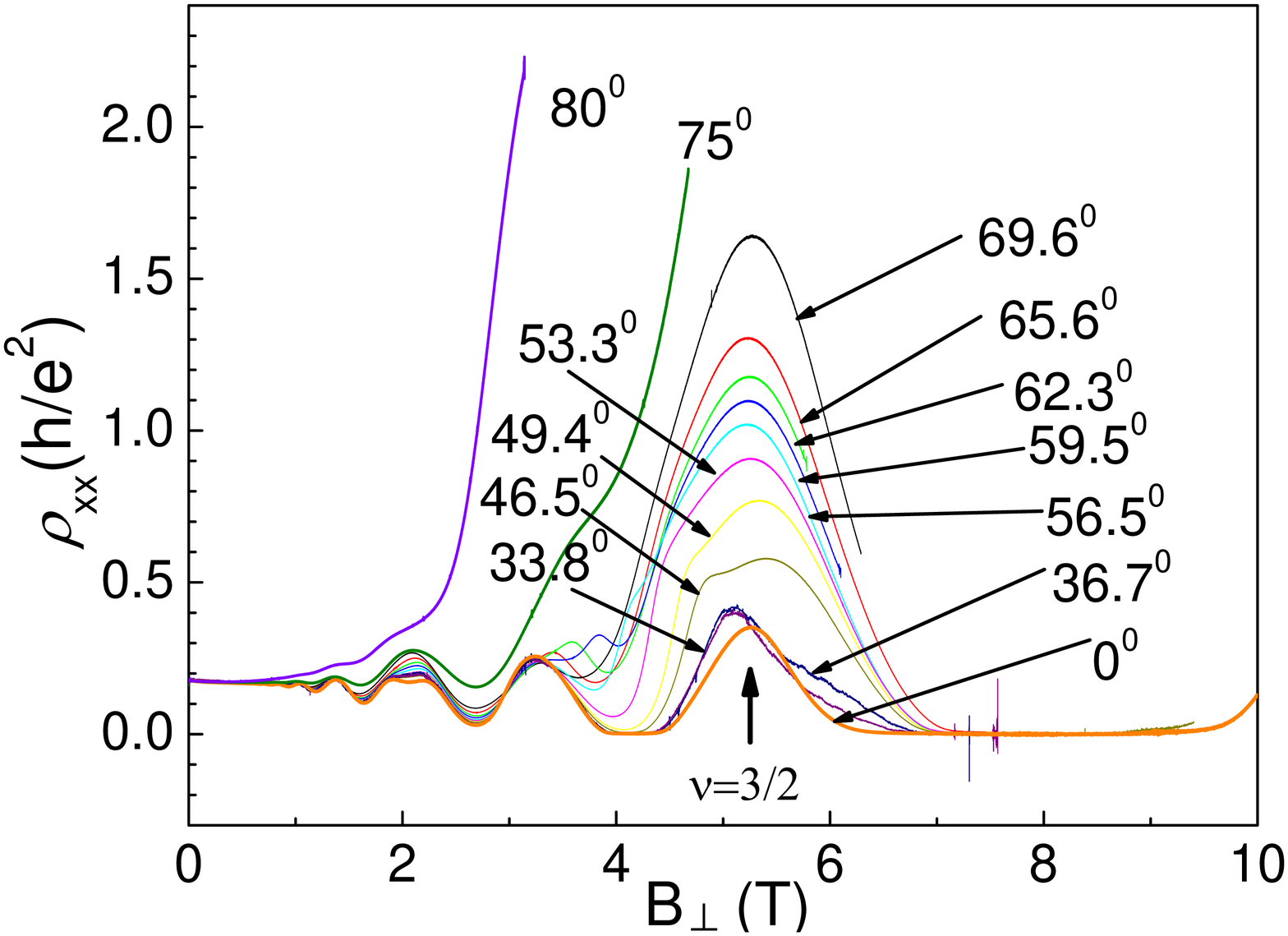}a)
\hfil
\includegraphics[width=7.8 cm,clip=]{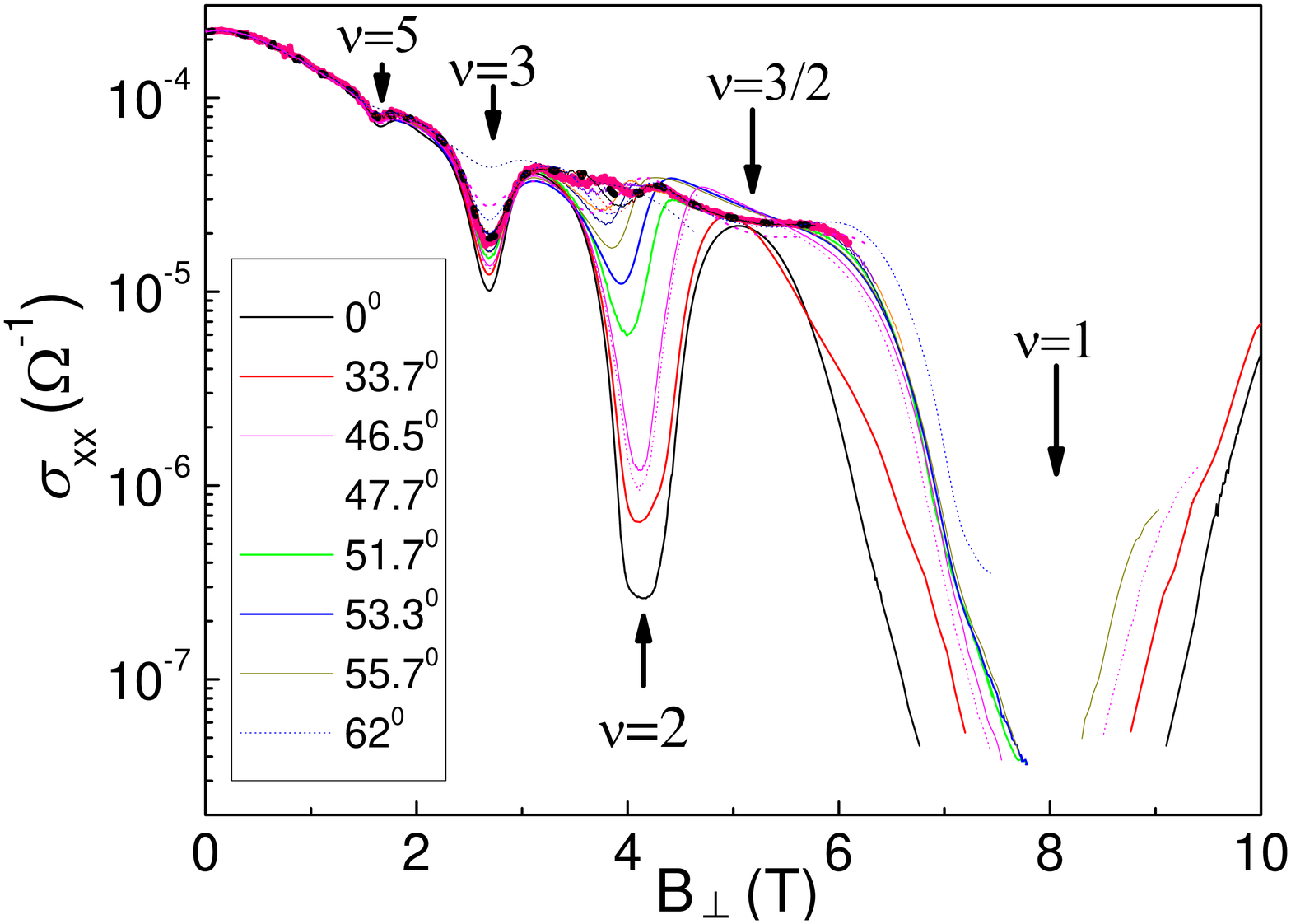}b)
\caption{Dependence of $\rho_{xx}$ (a) and $\sigma_{xx}$ (b) on $B_{\perp}$
at different tilt angles $\Theta$, T=0.3 K.}
\label{fig:2}
\end{figure}

Magnetoresistance and Hall effect were measured in tilted magnetic
field of up to 18 T in the temperature range of 0.3-1.6 K in the
linear regime at $I$ = 10 nA. More detailed studies of the tilt
effects were done on the sample with the density
$p$=2$\times$10$^{11}$\,cm$^{-2}$ and mobility
$\mu$=7$\times$10$^3$\,cm$^2$/Vs. These data allowed to calculate
the dependence of $\sigma_{xx}=\rho_{xx}/(\rho_{xx}^2+\rho_{xy}^2)$
on the magnetic field at different angles $\Theta$ between the 2D
interface surface normal and the magnetic field orientation
($\Theta$=0$^0$ if $B$ is perpendicular to the 2D interface). In
Fig.\ref{fig:2}, $\rho_{xx}$ and $\sigma_{xx}$ traces are plotted
versus normal component $B_{\perp}$ for a number of the tilt angles.

As can be seen from the Fig.\ref{fig:2} the maximum of $\rho_{xx}$
at $\nu$=3/2 increases with the increase of the tilt angle, which
was interpreted by many researchers as a M-I reentrant phase
transition. However, the dependence of $\sigma_{xx}$ on the angle at
$\nu$=3/2 does not show any anomaly. Further in this work the
analysis of the dependences of $\sigma_{xx}$ on the magnetic field,
temperature and the tilt angle will be presented.
\begin{figure}[t]
\includegraphics[width=6 cm,clip=]{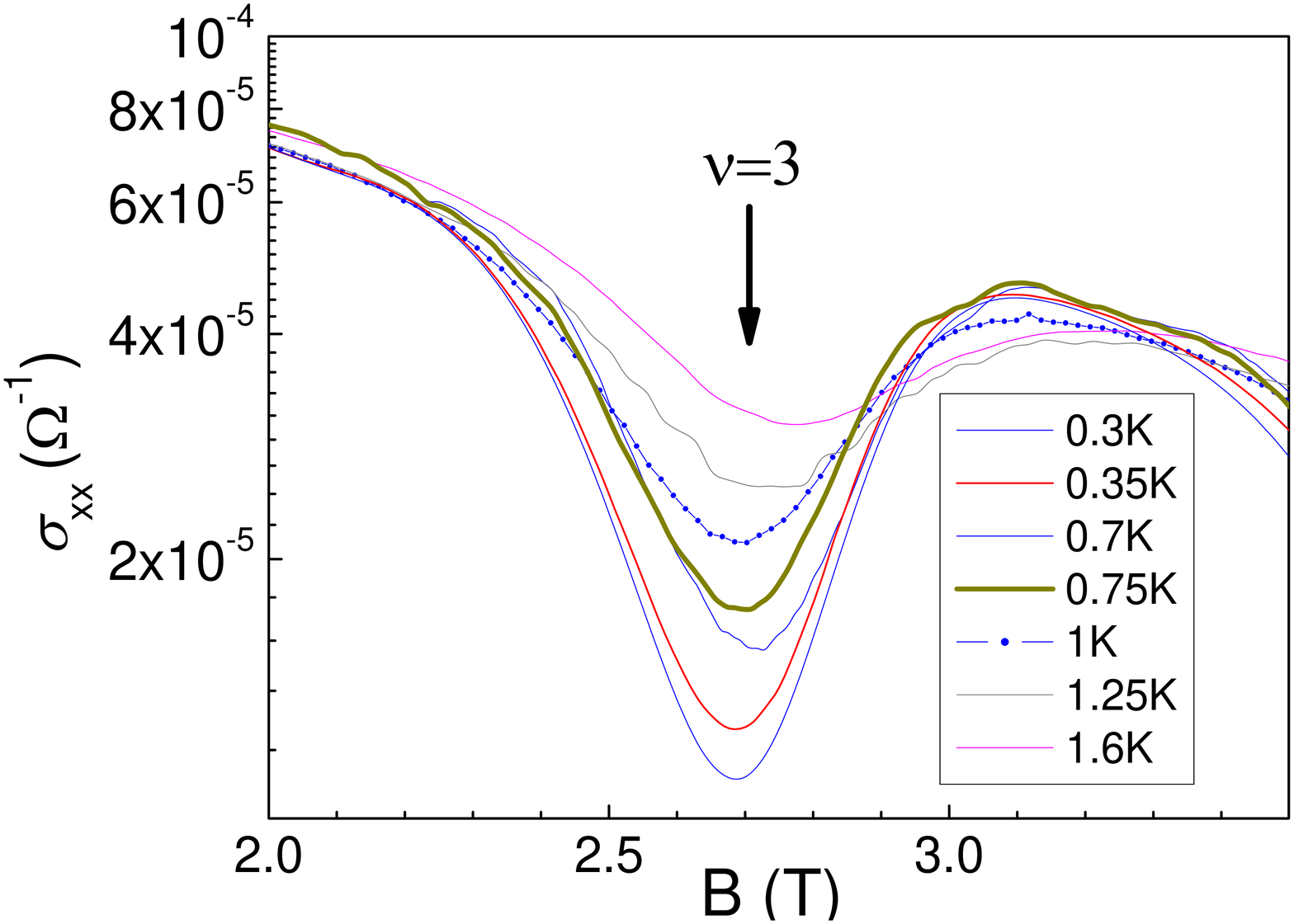}a)
\hfil
\includegraphics[width=6 cm,clip=]{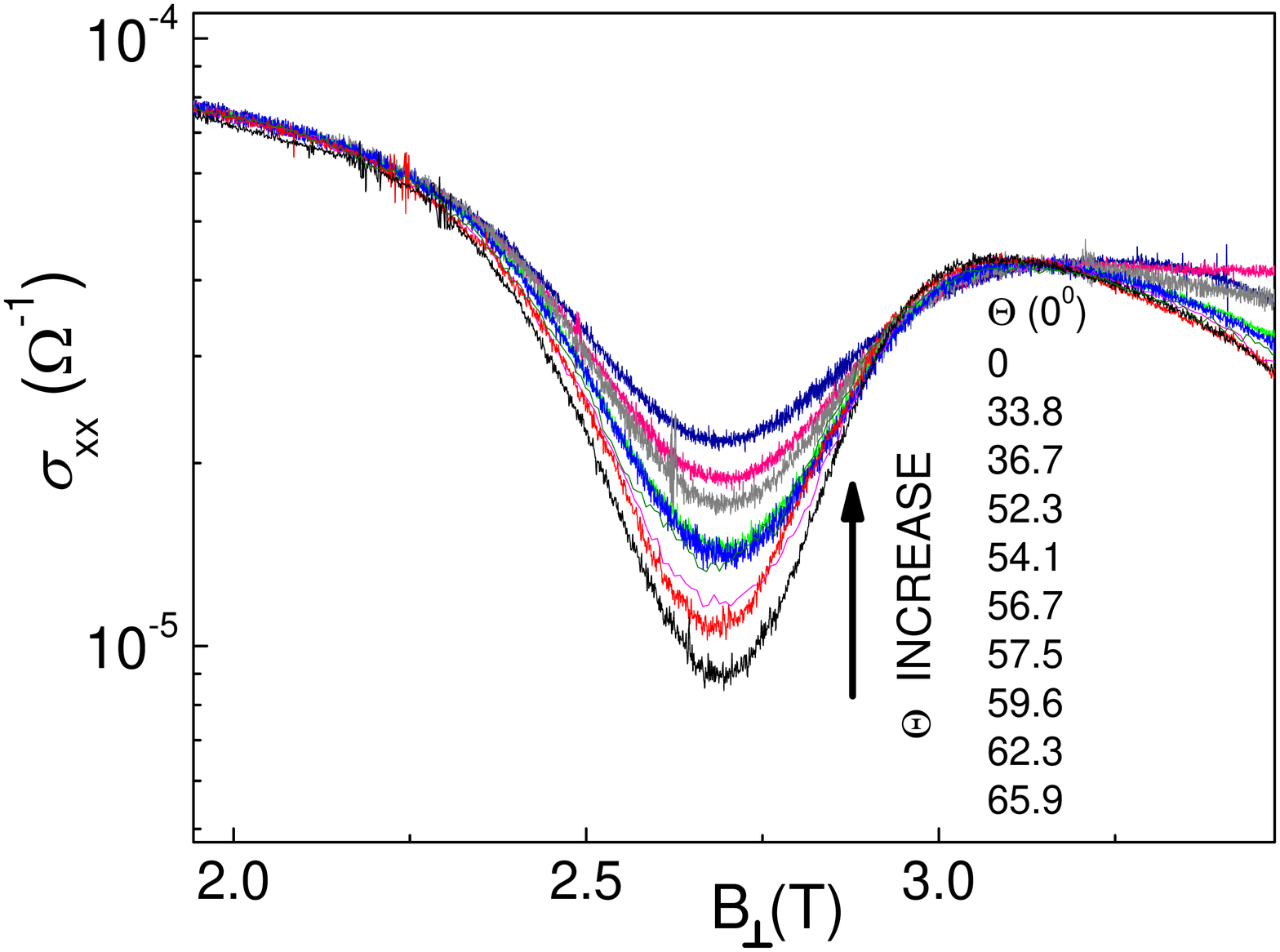}b)
\caption{(a) $\sigma_{xx}$ on $B_{\perp}$ at different temperatures,
$\Theta$=0, $\nu$=3; (b) $\sigma_{xx}$ vs $B_{\perp}$ for
various tilt angles at $T$=0.3K.}
\label{fig:3}
\end{figure}
\begin{figure}[h]
\includegraphics[width=60mm,height=40mm]{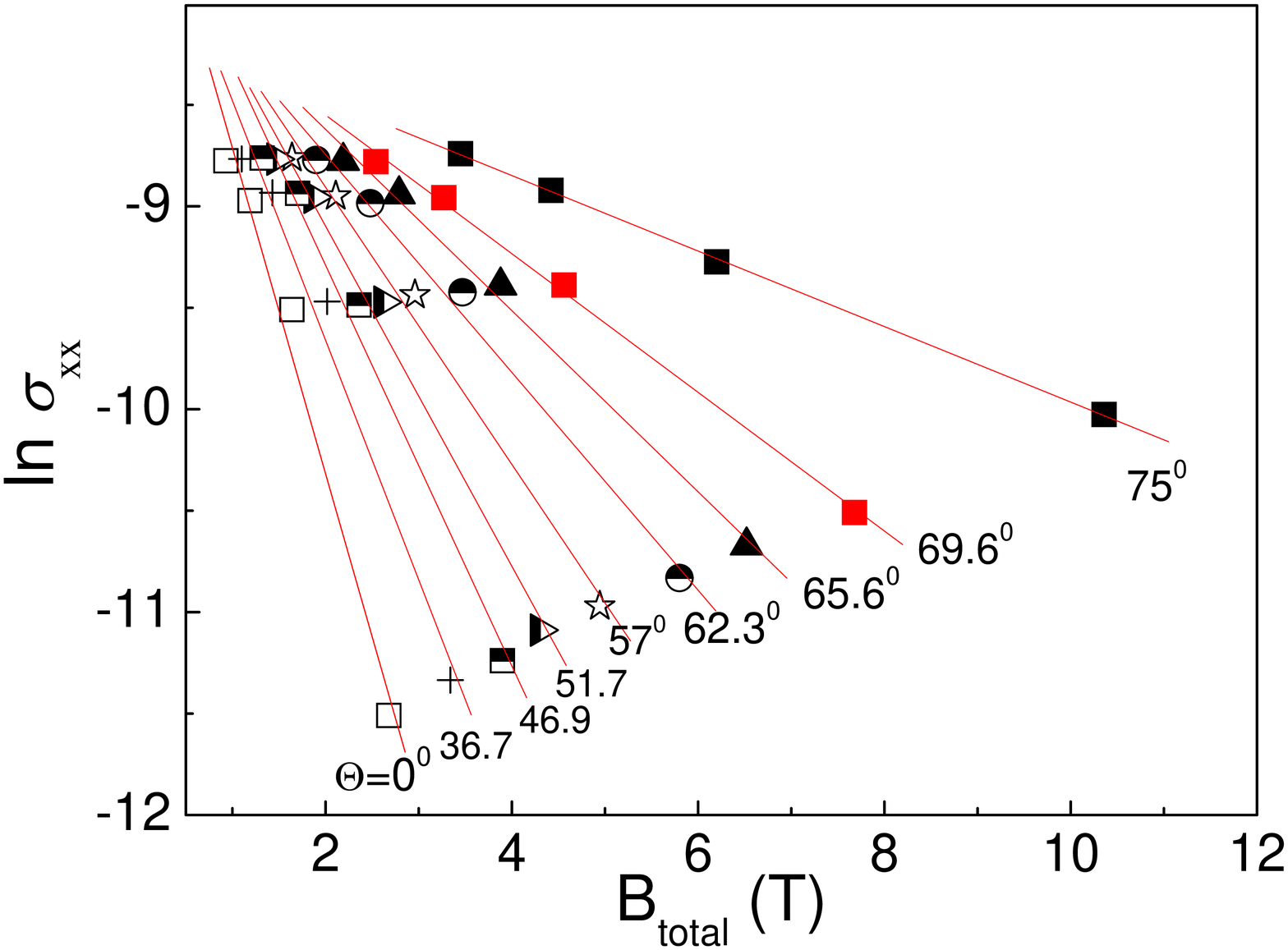}a)
\hfil
\includegraphics[width=60mm,height=40mm]{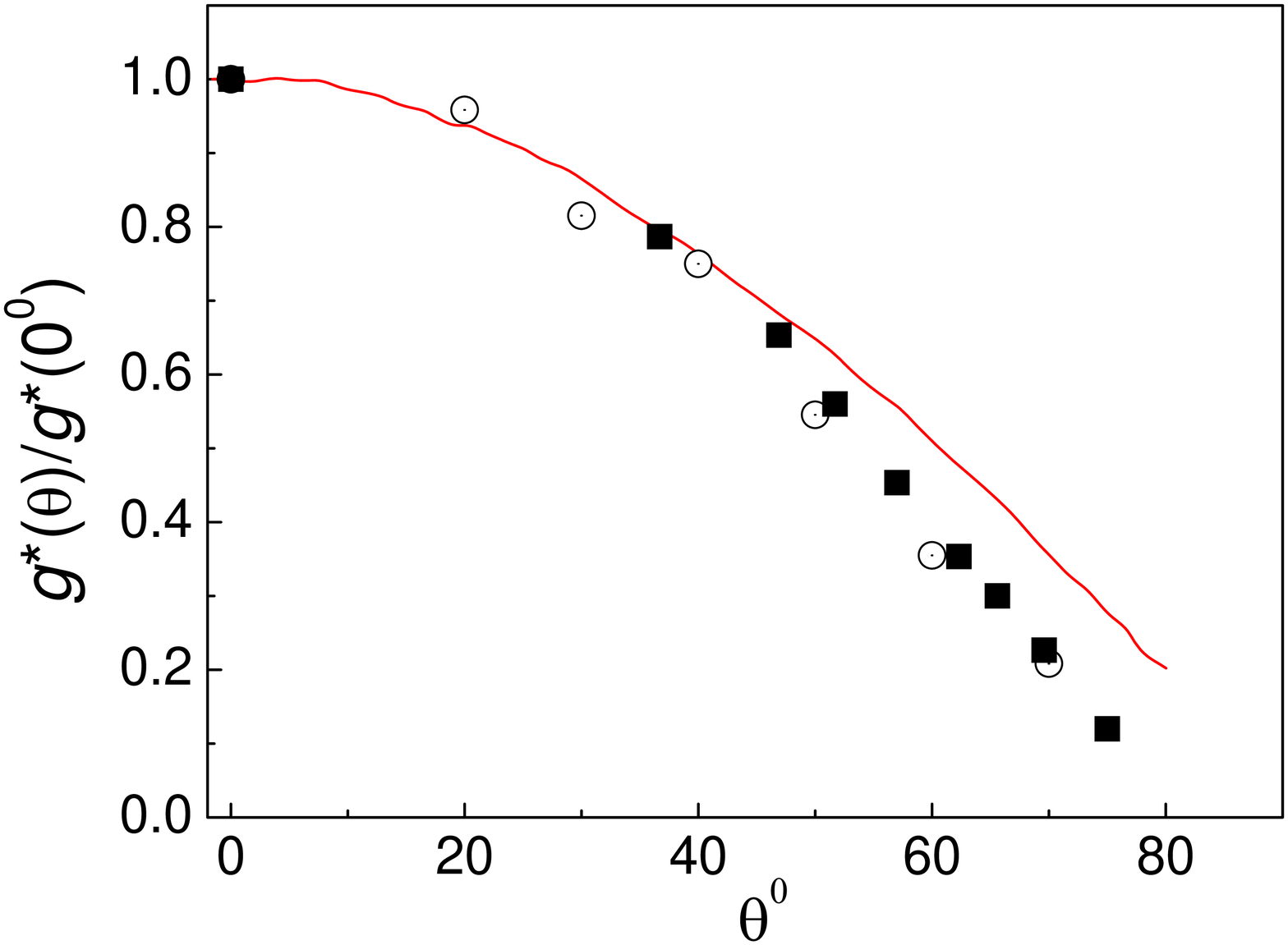}b)
\caption{(a) $\ln \sigma_{xx}^{min}$ vs $B$, $T$=0.3 K;
(b) reduced g-factor on tilt angle $\Theta$:
circles represents results of the angle-temperature association method,
squares - second method;  line is the g-factor from\cite{1}.}
\label{fig:4}
\end{figure}
As can be seen from the Fig.\ref{fig:3}, $\sigma_{xx}$ at $\nu$=3
increases with increasing temperature ($\Theta$=0$^0$) (a) as well
as with increase of the tilt angle ($T$=0.3 K = const) (b). Minimum
of $\sigma_{xx}$ at $\nu$=3 is observed when the Fermi level lies
between two spin split Landau levels 1$\uparrow$ and 1$\downarrow$.
In the temperature range of 0.6-1.7 K  $\sigma_{xx} \propto
\exp[-g^* \mu_B B/2k_BT]$, where $g^*$ is the effective g-factor,
$\mu_B$ is the Bohr magneton,  $B$ is the total magnetic field,
$k_B$ is the Boltzmann constant. %

The change of $\sigma_{xx}$ in Fig.\ref{fig:3}a is associated with
change of $T$ at constant g-factor and $B_{\perp}$. Yet the
conductivity variation with the tilt angle (see Fig.\ref{fig:3}b) at
constant $T$=0.3 is $\sigma_{xx} \propto \exp[-g^*(\Theta) \mu_B
B_{\perp}/2k_BT]$, and here this is g-factor that changes
(decreases). If to build the dependences $\sigma_{xx}$($T$) and
$\sigma_{xx}$($\Theta$) and at $\sigma_{xx}$ being equal to
associate the angle with a certain temperature $T'$ one may obtain
the equation $g^*(0^0) / 0.3 = g^* (\Theta) / T'$, i.e. $g^*(\Theta)
/ g^* (0^0) = T'/0.3$. Thus, one can determine the dependence of
reduced g-factor on the tilt angle.

\begin{figure}[t]
\includegraphics[width=73mm,height=49mm]{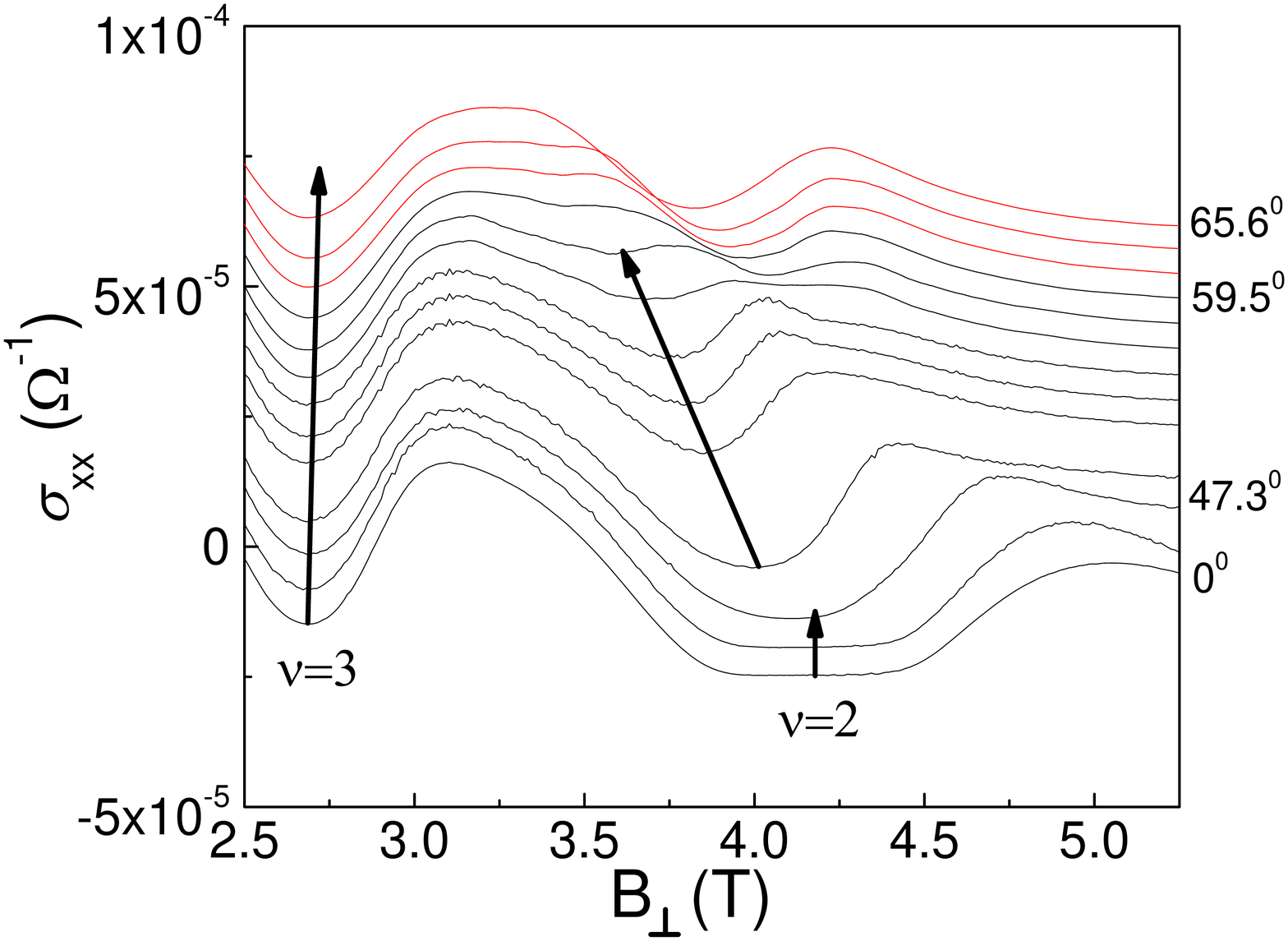}a)
\hfil
\includegraphics[width=68mm,height=45mm]{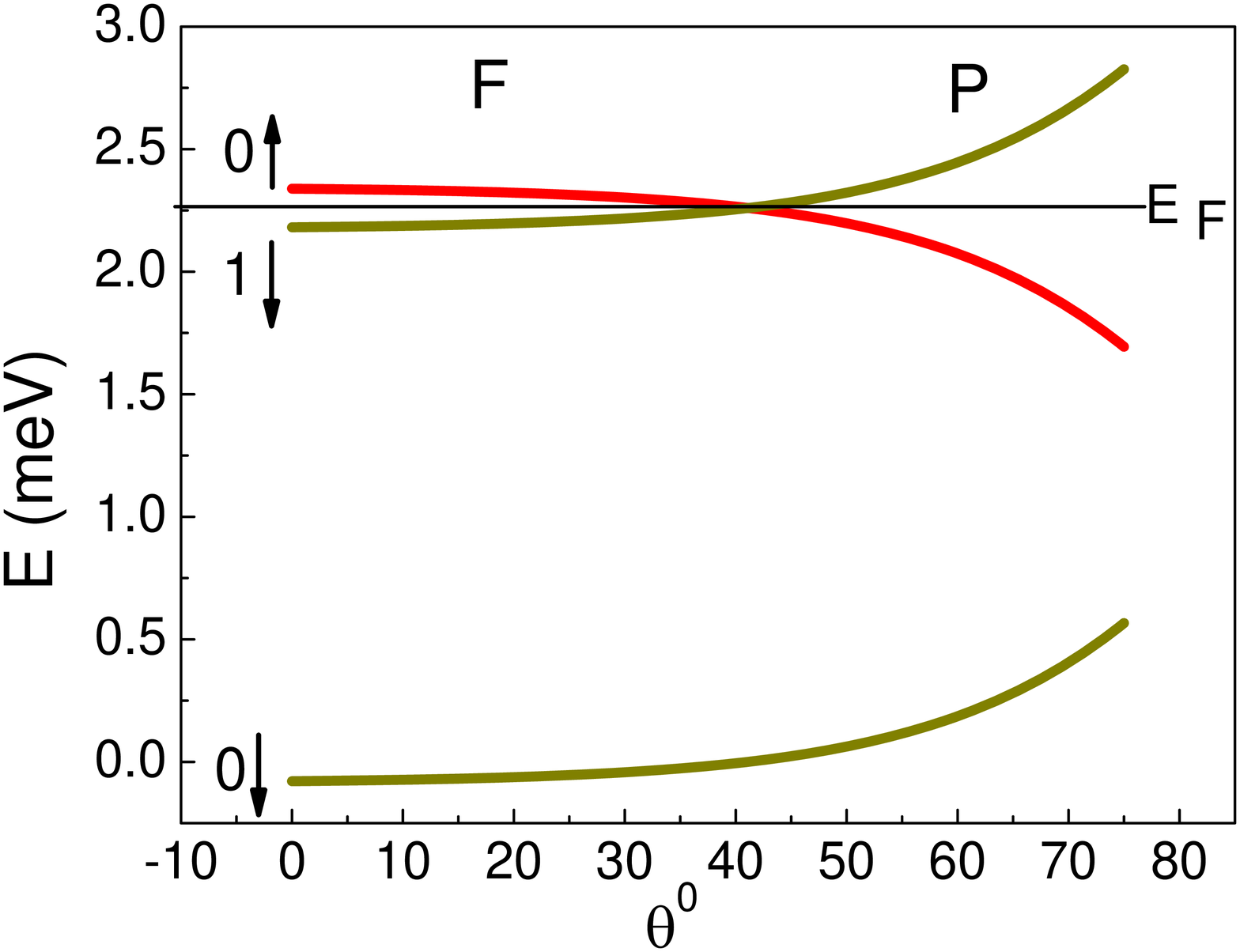}b)
\caption{(a) $\sigma_{xx}$ vs $B_{\perp}$ at the angles
$\Theta$=(0-65.6)$^0$, $T$=0.3 K. For clarity, the curves are
offset vertically by 5$\times$10$^{-6}$ohm$^{-1}$;
(b) Landau level energies vs. tilt angle:
at $\Theta$=0$^0$ there is a ferromagnetic state,
after crossing - paramagnetic.}
\label{fig:6}
\end{figure}

Another method for determining of $g^* (\Theta)$ is to construct
dependencies of $\sigma_{xx}$ magnitudes at $\nu$=3, 5, 7 and 9 on
total magnetic field $B$ at different angles $\Theta$
(Fig.\ref{fig:4}a). Since $\sigma_{xx} \propto \exp[-g^*(\Theta)
\mu_B B/2k_BT]$, then $\ln \sigma_{xx} (B) \propto g^* (\Theta)$ at
$T$=const and different $\Theta$, which makes it possible to
determine $g^*(\Theta) / g^*(0^0)$. Results of determination of
g-factor using these methods are illustrated in Fig.\ref{fig:4}b.
$\sigma_{xx}$ obeys to the activation law with $\Delta E/2k_B T
\approx 1.6$K. %
\begin{figure}
\includegraphics[width=7.8 cm,clip=]{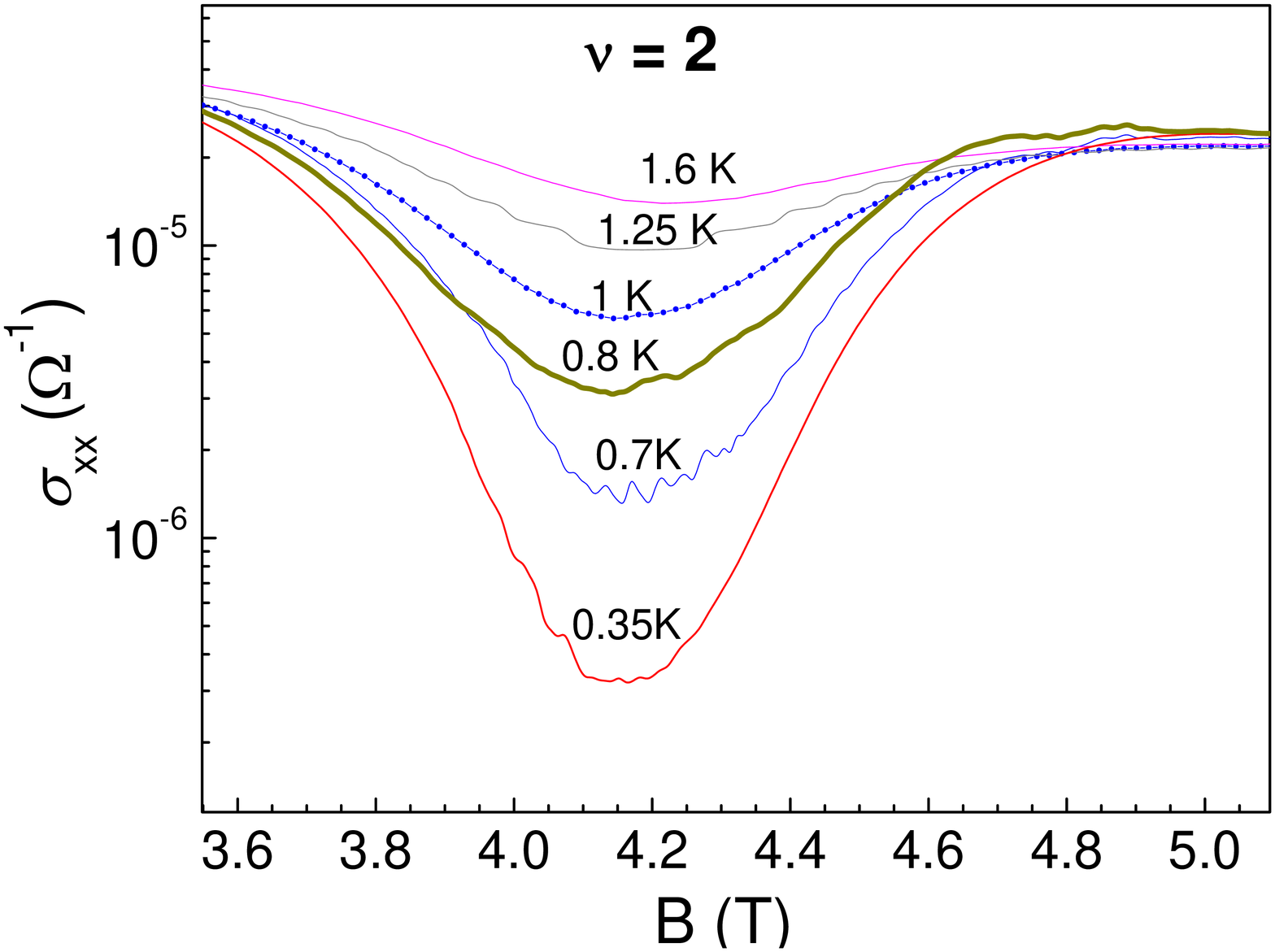}a)
\hfil
\includegraphics[width=7.8 cm,clip=]{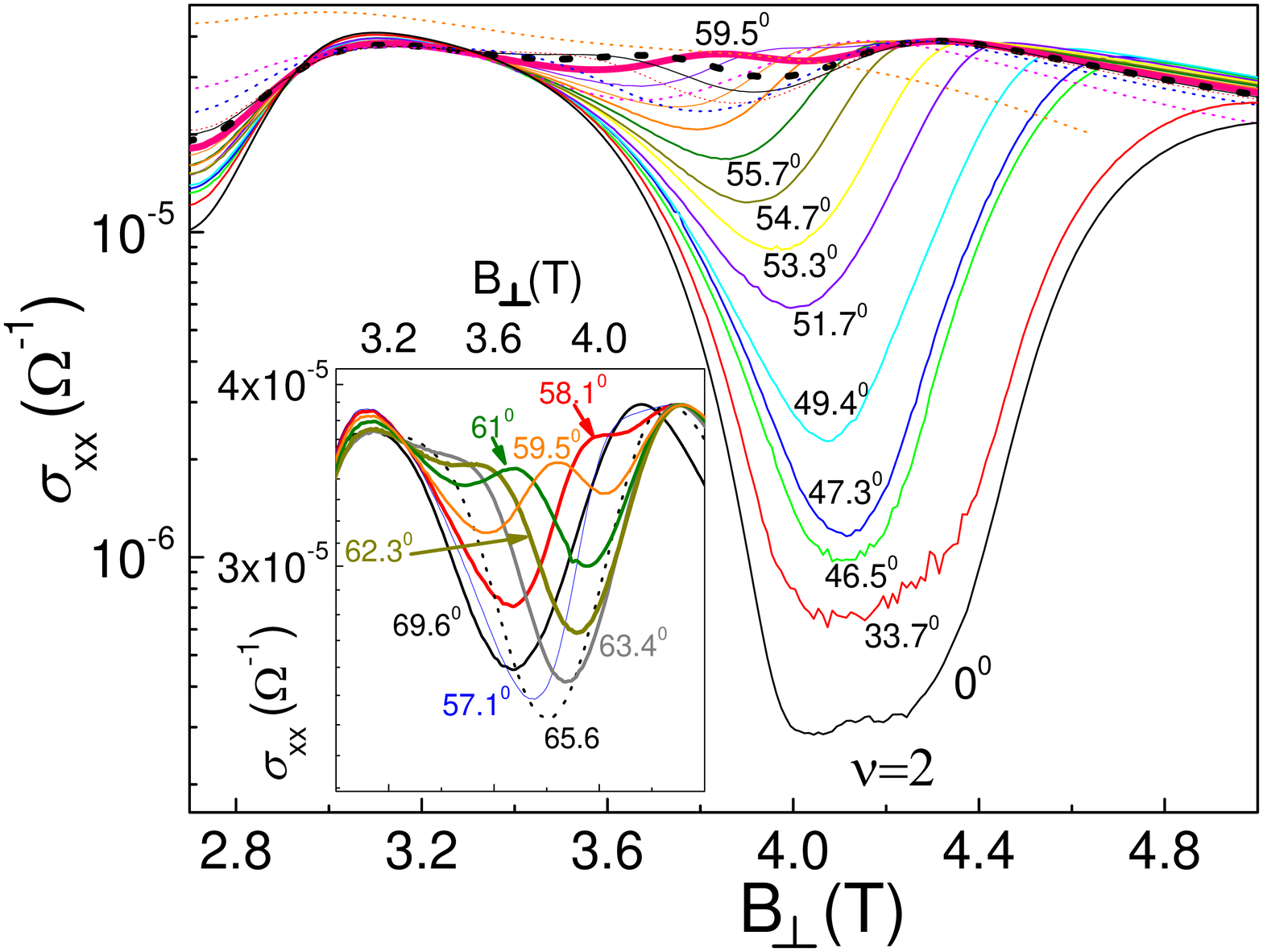}b)
\caption{(a) $\sigma_{xx}$ vs $B_{\perp}$ at different temperatures,
$\Theta$=0;
(b) $\sigma_{xx}$ vs $B_{\perp}$ at different tilt angles 0-65.6$^0$,
$T$=0.3 K, $\nu$=2. Inset: $\sigma_{xx}$ on $B_{\perp}$ at $\Theta$=(59.5-65.6)$^0$.}
\label{fig:5}
\end{figure}
We now consider changes of the conductivity oscillation traces near
$\nu$=2 with $T$ and $\Theta$. Fig.\ref{fig:5}a shows that the
position of the $\sigma_{xx}$ minimum at $\nu$=2 and $\Theta$=0$^0$
does not change in the magnetic field with temperature, and However,
with the increase of the tilt angle the conductivity minimum
increases and shifts in the direction of small magnetic fields until
$\Theta \approx$60$^0$ (Fig.\ref{fig:5}b). When the angle reaches
the value of $\approx$59.5$^0$, two oscillations appear on the
curve: the former, which shifted to the left with the angle
increase, and the new one which emerged at $B_\perp \approx$4 T.
With further increase of the angle this new oscillation shifts left
and grows in amplitude while the former oscillation disappears.
Fig.\ref{fig:6}a shows the new oscillation arising and a region of
angles where both types of oscillations coexist. The explanation of
this anomaly might be associated with the emergence of a
ferromagnetic-paramagnetic transition due to the Landau levels
crossing with the magnetic field tilt.
Fig.\ref{fig:6}b demonstrates the possibility of such crossing. We
used here the experimental data obtained in this work: the
dependence of $g^*(\Theta)/g^*(0^0)$ on $\Theta$, $\Delta E$=0.14
meV is the gap between the levels 0$\uparrow$ and 1$\downarrow$ at
$\Theta$=0$^0$. To enable Landau levels overlapping at $\Theta
\approx$(50-60)$^0$ it is necessary that at $\Theta$=0$^0$ the
system is in the ferromagnetic state, i.e. the energy of level
1$\downarrow$ is higher than the one of 0$\uparrow$, otherwise no
crossing occurs (in this case $g^* (0)=5$). It  is worth noting that
the results in Fig.\ref{fig:6}b are approximate, as a broadening of
the levels (disorder in the system) was not taken into account.

\subsection{Acknowledgments}

I.L.D. and I.Y.S. are grateful to Y.M. Galperin, L. Golub, S.
Tarasenko for useful discussions. This work was supported by grants
of RFBR 08-02-00852; the Presidium of the Russian Academy of
Science, the Program of Branch of Physical Sciences of RAS
"Spintronika". NHMFL is supported by the NSF through Cooperative
Agreement No. DMR-0084173, the State of Florida, and the DOE.

\end{document}